\documentclass[sigplan,10pt,screen]{acmart}

\AtBeginDocument{
  }

\usepackage[ruled,vlined,linesnumbered]{algorithm2e}

\usepackage{tikz}
\usepackage{amsmath}
\usepackage{pdfrender}

\usepackage{times,url,color,soul,xspace,enumitem}
\usepackage{graphicx}
\usepackage{subfig}
\usepackage{caption}
\usepackage{subcaption}
\usepackage{comment}
\usepackage{xspace}
\usepackage{booktabs}
\usepackage[inline,draft,nomargin,index]{fixme}

\usepackage{multirow}
\usepackage{cleveref}

\usepackage{pifont}

\usepackage{enumitem}

\acmSubmissionID{73}

\renewcommand\footnotetextcopyrightpermission[1]{}
\begin{document}

\newcommand{\GMG}{GMG\xspace}
\newcommand{\sys}{SliceScheduler\xspace}

\title
[\sys: Global Simulation-Guided Dynamic Operator Scheduling]
{
  Global Simulation-Guided Dynamic Operator Scheduling for Efficient Multi-Tenant Model Serving
}





\author{Weinan Liu}
\email{wnliu@stu.xmu.edu.cn}
\affiliation{%
  \institution{Xiamen University, Shanghai Innovation Institute}
  \city{Xiamen}
  \state{Fujian}
  \country{China}
}

\author{Zeyuan Ding}
\email{boris1013@sjtu.edu.cn}
\affiliation{%
  \institution{Shanghai Jiao Tong University}
  \city{Shanghai}
  \country{China}
}

\author{Dian Ding}
\email{dingdian94@sjtu.edu.cn}
\affiliation{%
  \institution{Shanghai Jiao Tong University}
  \city{Shanghai}
  \country{China}
}

\author{Chengcheng Wan}
\email{ccwan@sei.ecnu.edu.cn}
\affiliation{%
  \institution{East China Normal University, Shanghai Innovation Institute}
  \city{Shanghai}
  \country{China}
}

\author{Lu Tang}
\email{tanglu@xmu.edu.cn}
\affiliation{%
  \institution{Xiamen University}
  \city{Xiamen}
  \state{Fujian}
  \country{China}
}

\author{Guangtao Xue}
\email{gt_xue@sjtu.edu.cn}
\affiliation{%
  \institution{Shanghai Jiao Tong University}
  \city{Shanghai}
  \country{China}
}

\author{Jiwu Shu}
\email{shujw@tsinghua.edu.cn}
\affiliation{%
  \institution{Tsinghua University}
  \city{Beijing}
  \country{China}
}

\author{Yiming Zhang}
\email{sdiris@gmail.com}
\affiliation{%
  \institution{Xiamen University}
  \city{Xiamen}
  \state{Fujian}
  \country{China}
}

\newcommand{\method}[1]{\textit{\textbf{#1}}\xspace}
\newcommand*{\boldcheckmark}{%
  \textpdfrender{
    TextRenderingMode=FillStroke,
    LineWidth=.5pt, 
  }{\checkmark}%
}

\begin{abstract}

Container-granularity scheduling leaves abundant short-lived idle slices within containers unexploited.
Reallocating containers is too heavyweight to utilize such fine-grained opportunities under SLA constraints, and operator-level scheduling requires reasoning about dependencies, memory safety, and cluster-wide execution dynamics in real time. 

In this paper, we present \sys, a dynamic operator-level scheduling system for multi-tenant model serving.
The key idea is to expose cluster-wide operator execution state and enable what-if reasoning over scheduling decisions.
\sys consists of four key components.
First, we introduce the Global Mapping Graph (GMG),
a unified abstraction that captures operator dependencies, tensor shapes, resource mappings, and execution states,
providing a real-time, cluster-wide view with explicit resource semantics.
Second, we build a global simulator on top of GMG to predict operator-level execution and memory evolution under candidate placements.
Third, we design an incremental, simulation-based scheduling module that selects placements to exploit fragmented idle slices while avoiding memory violations and preserving SLA.
Finally, we develop an operator executor that materializes scheduling decisions on GPUs and coordinates computation and cross-accelerator transfers.
We implement \sys as a PyTorch backend and evaluate it using production trace replay.
Experimental results show that \sys improves token throughput by 1.10--2.29$\times$ compared to existing approaches,
while maintaining SLA violations within 9\%. \sys demonstrates that operator-level scheduling is a practical and effective approach to improving GPU utilization for multi-tenant LLM serving.

\end{abstract}


\maketitle

\section{Introduction}

To serve concurrent LLM inference requests from numerous users efficiently, 
large cloud companies typically 
provide GPU resources in a centralized way~\cite{acme, helios, philly, pai, optimus}.
As the cost of AI infrastructures grows substantially,
researchers have proposed various techniques to improve GPU utilization for LLM inference service (\emph{i.e.}, model serving) in multi-tenant clouds
\cite{optimus, pathways, alpa}.
%
The current mainstream practice is to perform tenant-granularity scheduling \cite{themis,sia} on top of container orchestration systems such as Kubernetes~\cite{k8s}.
Each LLM inference service is packaged into a container 
and submitted to a centralized scheduler,
which can allocate GPU resources to the container 
and reallocate them on demand,
\emph{e.g.}, when the containerized inference service has less requests during a long period of time. 

\begin{figure}[t]
\centering
\includegraphics[width=0.8\linewidth]{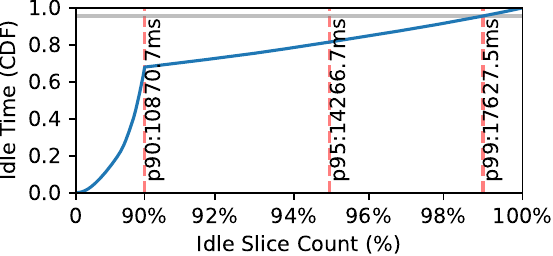}

    \vspace{-10pt}
\caption{
CDF of idle-slice durations from replaying a production AI coding traces on a single-GPU model instances.
Only idle slices longer than 10~ms and less than 20~s are included.
Average GPU utilization is 64.0\%,
and idle slices shorter than p99 account for 95.6\% of total idle time.
Idle slices are mostly short-lived.
}
\label{trace_replay}
\end{figure}


In modern Model-as-a-Service (MaaS) clouds,
GPU resources are often underutilized
\cite{pai}.
This is primarily because
tenant-granularity scheduling of containers 
does not align with the characteristics of model serving workloads.
To better understand this inefficiency,
we analyze the workloads of our production model serving clusters.
As shown in Figure~\ref{trace_replay},
there are a large number of short, fine-grained idle slices
(p50: 2551.11~ms, p90: 10887.64~ms, p99: 17790.31~ms),
as the inference services do not operate at full load at all times.
Unfortunately, 
container-level scheduling cannot utilize the idle slices 
due to strict service-level agreement (SLA) requirements of model serving.


The key problem is that
containers were originally designed for deployment~\cite{k8s1} rather than for scheduling.
Once the scheduler allocates GPUs to an inference service container,
the container drives GPU execution,
while the scheduler no longer has direct control over it.
As a result,
the idle slices within a container’s execution
logically belong to that container,
and the scheduler cannot directly reallocate them to others.
%
Container resource reallocation requires preemption and rescheduling,
which typically involves checkpointing, re-queuing (under gang scheduling~\cite{gangscheduling}), runtime cold start, and state restoration.
This overhead is too high to operate at the time scale of second-level idle slices.
Essentially,
the container abstraction isolates the scheduler from GPU execution, 
limiting its visibility into and control over the numerous short-lived idle slices across the cluster.

To allow the scheduler to directly control GPU execution,
scheduling must move below the container level to 
finer-grained execution units in inference services.
Operators,
such as kernels, ATen~\cite{aten} ops, and LLM layers,
readily fit into short-lived idle slices,
making them natural scheduling units.
This enables an inference service to no longer be tied to a fixed set of containers/GPUs;
instead,
its execution can be spread across the cluster.

Although operator-level scheduling has been widely adopted in 
prior graph-based job schedulers
\cite{mapreduce,dryad,spark,flink,starpu}, 
where jobs are represented as graphs and nodes (operators) are mapped to resources,
applying operator-level scheduling to multi-tenant model serving remains challenging. 
To achieve this,
the scheduler must track short-lived, fragmented idle slices in real time during inference execution, 
and schedule operators into them without violating the SLA.
This brings the following three challenges.

\textbf{Challenge 1: Lack of operator resource requirement undermines memory safety.}
Prior graph-based schedulers model dependencies among graph nodes,
but do not capture the operator-level resource requirement of individual operators.
Specifically, 
they do not explicitly model the input/output memory footprint of each operator, 
which can cause an operator placement to trigger
HBM out-of-memory (OOM). 
Dependency information alone is insufficient to ensure memory-safe scheduling on fragmented GPU resources.

\textbf{Challenge 2: Lack of temporal execution prediction 
limits scheduling effectiveness.}
Existing graph-based schedulers map graph nodes to resources to optimize specific objectives (\emph{e.g.}, data locality),
but lack the capability to predict how inference execution evolves over time at cluster scale.
Static operator-resource mapping is insufficient,
as 
short-lived idle slices emerge dynamically between operator executions 
and require the scheduler to 
reason not only about static operator placement 
but also about evolving execution states.

This paper presents \emph{\sys},
an efficient operator-level scheduler 
that dynamically schedules operators (\emph{layers}) into idle slices 
to improve GPU utilization without violating SLA constraints.
The key idea of \sys is to 
expose cluster-wide, fine-grained operator execution state to the scheduler
so that scheduling decisions can be simulated and assessed 
before being applied in a real multi-tenant model serving cluster.
\sys address the aforementioned challenges with the following designs.


First,
we propose \emph{Global Mapping Graph} (\GMG),
a cluster-wide representation of active operators
with their dependencies, tensor shapes, and resource mappings.
Since each operator's tensor shape can be inferred before execution,
operators in the \GMG have explicit resource requirements.
This allows the \GMG to capture HBM occupancy across the cluster,
together with tensor residency and operator dependency.
Operators are inserted to and removed from the \GMG over time,
providing a real-time view of the cluster state.

Second,
we design a global simulator on top of \GMG
that predicts how operator executions evolve across the cluster under a given scheduling decision.
Operator execution is predictable, 
as operator latency is largely determined by compute-intensive kernel implementations with stable execution times for a given shape \cite{hap}.
Similarly, tensor transfer is also predictable. 
Leveraging this predictability,
together with the dependencies and resource mappings in \GMG,
we can precisely simulate future cluster states.

Third,
based on \GMG and global simulation,
we propose dynamic operator-level scheduling for multi-tenant inference services.
Users submit inference requests as \emph{computation graph} instances with priority, 
which are inserted into the \GMG.
\sys plans and evaluates candidate operator scheduling policies via parallel what-if simulations,
selecting the one that avoids OOM and achieves the best predicted completion time and resource efficiency.

Finally,
we design an operator executor to materialize selected scheduling decisions on GPUs.
We first consider single-GPU execution by introducing an operator queue,
and then extend to cross-GPU tensor transfers to explicitly model data movement between GPUs.
Tensor transfers are managed by an arbitration algorithm,
which enables \sys to coordinate distributed computation and communication under a unified runtime.
Together,
these mechanisms provide the execution foundation for cluster-wide operator scheduling.

We evaluate \sys by replaying production traces.
The results show that \sys can opportunistically insert offline workloads into second-level idle slices left by online workloads,
while achieving load balancing across the cluster.
Compared with the online-only and GPU-sharing baselines,
\sys improves token throughput by 1.10$\times$--2.29$\times$ 
and Service Level Objective (SLO) attainment by $\sim$9\%.
We further show that dynamic operator-level scheduling is practical:
the global simulator sustains an average throughput of $\sim$176 simulated operators per millisecond,
achieving about four orders of magnitude higher throughput than prior approaches~\cite{simai}.


We make following contributions:

\vspace{-1mm}
\begin{itemize}
\item
We introduce \GMG,
an operator-level abstraction for multi-tenant model serving 
that enables global simulation of operator execution and HBM occupancy.

\item
We design \sys,
a dynamic operator-level scheduler for multi-tenant model serving,
and propose a simulation-based scheduling algorithm.

\item
We implement \sys as a PyTorch~\cite{pytorch} backend, 
and validate the effectiveness of dynamic operator scheduling on real-world workloads.

\end{itemize}

\section{Background \& Motivation}


Modern multi-tenant model serving clusters mainly adopt container-level scheduling~\cite{k8s,acme,philly,pai}.
However, production traces show that 
there are many short-lived idle slices that such scheduling fail to utilize.
This section first characterizes these idle patterns 
and then explains the main challenges in utilizing them.

\subsection{Trace Characterization}

\label{sec:trace}


\begin{figure}[t]
\centering
\subfloat[Request volume, aggregated over 10-minute intervals.]{\includegraphics[width=1\linewidth]{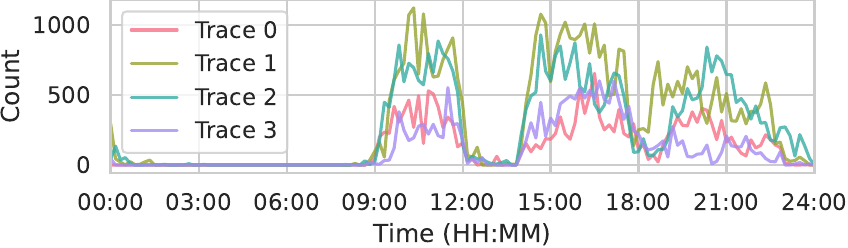}}
\newline
\subfloat[Conservative model-instance computation time in each 10-minute interval.]{
\includegraphics[width=1\linewidth]{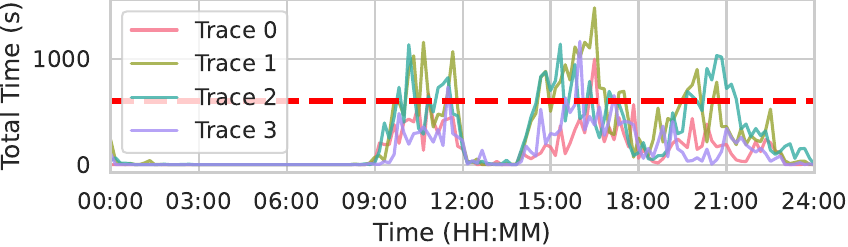}}

    \vspace{-9pt}
\caption{
Four one-day production traces from an AI-coding service. 
}
    \label{motivation23}
\end{figure}



We collect two-week AI coding traces from an industrial company.
Each request record includes its arrival timestamp, context length, output length, time to first token (TTFT), and time per output token (TPOT).
From this trace,
we select four representative days and analyze them as four one-day traces.
As shown in Figure~\ref{motivation23}~(a),
the workload exhibits clear diurnal patterns.
All four traces share same peak periods during 08:00--12:30, 14:00--18:00, and 19:00--23:59.

Using TTFT and TPOT of the traces,
we conservatively compute the model-instance compute time within each 10-minute window,
as shown in Figure~\ref{motivation23}~(b).
Most windows have compute time below 600~s (\emph{i.e.}, window duration),
indicating that the model instance is far from fully utilized in most times.
Even during peak periods,
the demand remains uneven across traces:
while one trace may impose more than 600~s of aggregate compute demand in a window,
another may still remain below it.
This means that,
at the same time, some instances can be overloaded while others are underutilized,
revealing opportunities for cross-instance load balancing.


To further understand the headroom,
we replay the trace on a Llama3 8B~\cite{llama3} instances deployed on one A16 GPU.
As shown in Figure~\ref{trace_replay},
the average GPU utilization is 64.0\%,
and the execution contains a large number of idle slices.
Their durations have a median of 2551.11~ms, a 90th percentile of 10887.64~ms, and a 99th percentile of 17790.31~ms.
Idle slices shorter than the 99th-percentile duration account for 95.6\% of total idle time,
showing that idle slices are substantial in aggregate.
This provides performance enhancement opportunity by utilizing such resource fragmentation.

\begin{figure}[t]
\centering
\subfloat[Execution timeline of a high-priority online instance during trace replay (blank slices refers to idle).]{\includegraphics[width=0.85\linewidth]{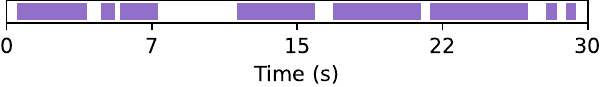}}
\\
\subfloat[
Requests from a low-priority offline instance could utilize idle slices (yellow slices).
]{
\includegraphics[width=0.85\linewidth]{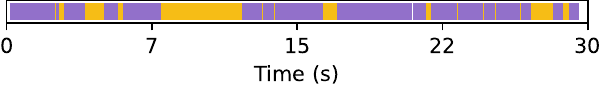}}

    \vspace{-9pt}
\caption{
An illustrative co-location example @ 1 GPU.
}
\label{op_motivation}
\end{figure}




Next, we conduct a small co-location experiment with two Llama3~8B~\cite{llama3} instances on a single A16 GPU:
a high-priority online instance and a low-priority offline instance.
During trace replay,
the online instance exhibits substantial idle slices.
We then opportunistically place requests from the offline instance into those gaps.
As shown in Figure~\ref{op_motivation},
this converts part of the online instance’s idle time into useful work.
With the SLA maintained,
the throughput of tokens is improved by $1.25\times$.
It suggests that fine-grained idle slices left by online services can in principle be reclaimed by low-priority work,
motivating a scheduling abstraction finer than containers.

\subsection{Why Container-Level Scheduling Cannot Exploit Fine-Grained Idle Slices}



The traces above show substantial fine-grained idle slices in online ML services.
Note that,
these idle slices does not simply reflect misprovisioning.
Services are often provisioned near saturation for SLA reasons,
yet still generate transient idle slices due to bursty demand.
Existing container-level schedulers fail to exploiting idle slices for two reasons: coarse-grained control and exclusive resource ownership.

\subsubsection{Coarse-Grained Control}

Container orchestration systems were originally designed for application deployment rather than direct control of accelerator execution~\cite{k8s1}.
A scheduler selects a node that satisfies a container’s resource request,
after which the target node pulls the image,
initializes the runtime environment,
launches the application process,
and lets the process drive the assigned GPUs.
In other words,
container schedulers control resource usage only indirectly through the application runtime,
rather than directly managing GPU compute units or HBM.

This design naturally binds a container to a fixed set of accelerators for a relatively long period after launch.
To handle longer-timescale load variation in ML workloads, prior systems introduce elasticity~\cite{pollux,lyra},
which reclaims resources through preemption and rescheduling.
But elasticity is expensive:
the application must first checkpoint,
and then go through rescheduling,
gang-scheduling~\cite{gangscheduling} delays,
image pulling,
environment initialization,
and checkpoint loading.
In practice,
the end-to-end overhead is often on the order of minutes.

As a result,
elasticity is useful mainly for long-timescale reallocation.
For example,
prior work uses it to search for better configurations for long-running training jobs~\cite{pollux},
or to exploit diurnal variation by shrinking online services at night and reallocating resources to other workloads~\cite{lyra}.
Such mechanisms are ill-suited to our scenario.
For online services with frequent second-scale idle slices,
simply reducing the number of GPUs directly hurts SLA attainment;
conversely,
using elasticity to let low-priority jobs opportunistically fill these idle slices is impractical because the rescheduling overhead is far too high.

At a deeper level,
the mismatch comes from the unit of control.
A container scheduler does not move a few kernels or a small region of HBM;
it migrates and reconstructs an entire application runtime and execution environment.
Therefore,
container-level control is fundamentally too heavyweight to track and exploit fine-grained execution gaps on accelerators.

\subsubsection{Exclusive Resource Ownership}

A second challenge is that these fine-grained idle slices still logically belong to the original containers.
Even when an online service does not fully utilize GPU compute or HBM at every moment,
these slices remain allocated to that container.
Since we cannot modify users’ applications,
we cannot expect them to voluntarily expose or coordinate these fine-grained resources for use by other workloads.

Prior works like GPU sharing~\cite{orion,antman} recognize that exclusive GPU ownership can lead to low utilization on a single device,
and thus seek to co-run multiple jobs on one GPU.
However,
such approaches still do not address the kind of fragmentation we face.
One line of GPU sharing partitions compute and HBM statically~\cite{mig},
reserving a fraction of each GPU for low-priority jobs.
This enables steady co-location,
but it is fundamentally a fixed spatial partitioning scheme:
it does not capture the transient idle slices generated by the high-priority workload during execution.

Another line allows multiple jobs to share a GPU without static HBM partitioning~\cite{mps,antman}.
Such approaches are most effective when co-located training jobs have overlapping but manageable memory-demand patterns.
In modern Large Language Model (LLM) serving,
however,
model weights and runtime state tend to stay resident in HBM,
making memory pressure, swapping, and performance interference much more likely during peak demand.
More importantly,
these methods remain poorly suited to exploiting idle slices that is fragmented across multiple GPUs and containers.

This is because online services typically occupy most of the HBM,
leaving only small residual memory regions scattered across different GPUs for lower-priority jobs.
To exploit such fragments,
the low-priority job itself must be decomposable into fine-grained pieces and spread across multiple GPUs,
potentially crossing container boundaries.

At the same time,
as cluster scale grows,
the total amount of idle slices may increase,
but idle slices on different GPUs rarely align in time.
To validate this,
we replay the four representative one-day traces mentioned above on four model instances.
Surprisingly,
no more than 20\% of the idle time overlaps across the four traces.
The more GPUs the low-priority job needs to cover,
the less likely they are to become idle simultaneously.
Consequently,
if a low-priority job is still deployed as a GPU-sharing container,
it often requires multiple GPUs to be concurrently available in order to make progress on its end-to-end computation.
This significantly limits the amount of idle slices it can exploit and increases the likelihood of interfering with existing online workloads.

Even if such methods can run a job on a small number of GPUs,
the job still cannot exploit idle slices on other GPUs in the cluster.


\subsection{Requirements for Exploiting Fine-Grained Idle Slices}

Exploiting fine-grained idle slices requires more than moving below containers.
The system must answer four questions.




\textbf{How to design resource semantics that capture operator-level hardware requirements?}
Operator shapes are sufficient for this purpose.
Given an operator's shape,
its execution time on a particular accelerator is largely predictable.
Likewise,
given tensor shapes,
transfer latency across accelerators is also predictable.
Therefore,
we uses the \GMG as a unified abstraction of cluster state,
and every operator inserted into \GMG carries explicit shape information.

\textbf{How to reconstruct the execution timeline and reason about cluster evolution?}
With operator dependencies,
the completion of one operator will trigger its successors,
causing cluster execution state to evolve over time.
Thus we design an event-driven global simulator that advances execution along operator dependencies.

\textbf{How to design the whole scheduling mechanism?}
\GMG serves as the core abstraction maintained by the master.
Serving applications express computation graph execution.
Then we build an independent scheduling module that works together with \GMG.
During scheduling,
the module constructs global simulators and uses them to generate operator-granularity scheduling policies.

\textbf{How to design an operator executor that realizes scheduled execution?}
We first design operator execution on a single accelerator,
and then extend it to tensor transfer across accelerators.
Finally,
we introduce the tensor transfer arbitration to coordinate tensor movement among all accelerators in the cluster.

Together,
these requirements point to a scheduling layer that is
(i) the \GMG for cluster-wide state abstraction with operator shape,
(ii) a global simulator,
(iii) a scheduling module, and
(iv) an operator executor.
The rest of this paper presents such a design.

\section{\sys Design}

To tackle these problems,
we propose \sys,
an efficient operator-level scheduler that can dynamically schedule low-priority operators into high-priority operators’ idle slices and significantly improve GPU utilization without violating SLA constraints.

\subsection{System Overview}

\begin{figure}[t]
    \centerline{\includegraphics[width=0.5\textwidth,]{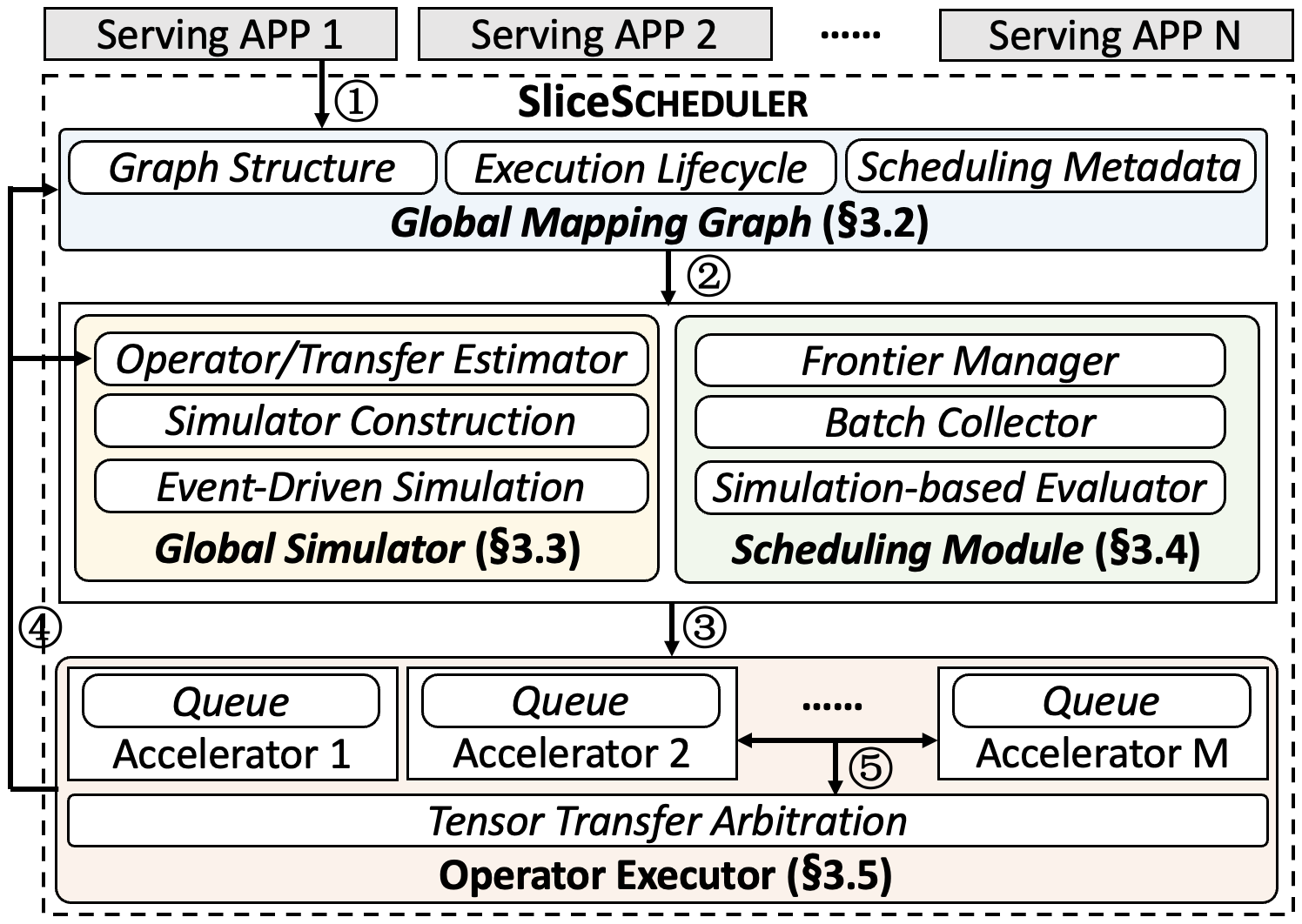}}
    
    \vspace{-5pt}
    \caption{
Overview of \sys.
The system centers on the Global Mapping Graph (GMG),
which serves as a unified abstraction of computation, placement, and runtime state.
Built on top of GMG, the Global Simulator predicts execution dynamics,
the Scheduling Module performs simulation-based placement decisions,
and the Operator Executor executes operators and transfers on accelerators.
Runtime feedback continuously updates GMG and estimators,
enabling a closed-loop, prediction-driven scheduling system.
    }
    \label{overview}
 
\end{figure}

\textbf{Overview.}
As shown in Fig.~\ref{overview}, \sys decomposes cluster-scale scheduling into four tightly coupled layers:
state abstraction, prediction, decision, and execution. 
At the core is \textbf{the Global Mapping Graph (\GMG)},
which provides a consistent cluster-wide view of operators, dependencies,
tensor shapes, and their placement on accelerators.
This global abstraction is necessary for making coherent scheduling decisions.
However, scheduling depends not only on the current state,
but also on its future evolution.
To this end, \sys builds \textbf{the Global Simulator} on top of \GMG,
which reconstructs execution behavior and predicts operator execution
and tensor transfer timelines under candidate placements.
Based on these predictions, \textbf{the Scheduling Module} selects placements
through incremental what-if simulation,
enabling fast yet globally informed decisions.
Finally, \textbf{the Operator Executor} realizes these decisions on accelerators,
while feeding runtime feedback back to \GMG and the estimators.

\sys consists of a centralized master and a set of accelerators that execute operators.
The master maintains the \GMG as a global state abstraction, where serving applications perform \textbf{(1) Registration and instantiation} by registering graph templates and instantiating graph instances whose operators are inserted as \textit{unscheduled}.
From this evolving state, \sys performs \textbf{(2) Scheduling and simulation} by selecting a batch of operators and invoking the scheduling module, which constructs global simulators to evaluate candidate placements via what-if simulation.
After selecting a placement, \sys proceeds with \textbf{(3) Issuing} by updating operator-to-accelerator mappings in \GMG, marking operators as \textit{issued}, and dispatching them together with required transfers to the executors.
During execution, accelerators process operators via dependency-driven queues while handling \textbf{(4) Transfer arbitration} to coordinate cross-accelerator tensor transfers.
As operators and transfers complete, \sys performs \textbf{(5) Feedback} by reporting runtime information back to the master, marking operators as \textit{done}, and updating performance estimators to improve future simulation and scheduling decisions.



\subsection{Global Mapping Graph}

\begin{figure}[t]
    \centerline{\includegraphics[width=0.5\textwidth,]{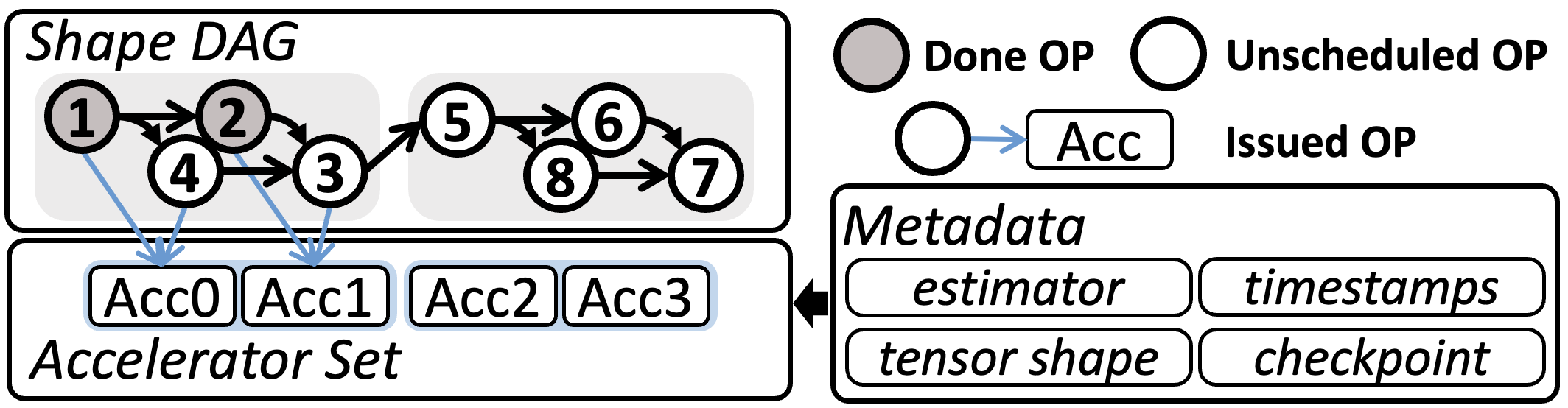}}
    
    \vspace{-5pt}
    \caption{\GMG maps operators to accelerators.
It describes the global status of tasks and the cluster,
as well as the computation progress of all tasks.
    }
    \label{GMG}

\end{figure}

Existing systems lack a unified abstraction that captures both computation structure and resource states at operator granularity, making it difficult to reason about scheduling decisions globally. \GMG unifies computation structure, placement, execution state,
and scheduling metadata into a single abstraction.

\textbf{Graph structure.}
The core of \sys is the Global Mapping Graph (\GMG).
As shown in Figure~\ref{GMG}, 
\GMG consists of a Shape DAG and an Accelerator Set.
In the Shape DAG,
each node represents an operator annotated with its tensor shape and executor state, and each edge represents a data dependency.
Operator nodes are mapped to accelerators in the Accelerator Set.
The executor state of an operator indicates whether it is
\textit{unscheduled},
\textit{issued},
or \textit{done}.

\textbf{Execution lifecycle.}
We design the computation pattern as a register-once, instantiate-many-times workflow.
A serving application first registers a graph template with \sys,
and then repeatedly instantiates this template for subsequent executions.
All graph instances derived from the same template share the same graph topology,
while the shapes of individual operators may vary across instances.
For each new execution instance,
the application specifies its inputs by referencing existing operators in \GMG.
The registered template contains symbolic shape information,
allowing \sys to infer the shape of every operator in the instance from the input shapes.
After shape inference,
the instantiated graph is inserted into \GMG as a set of \textit{unscheduled} operators.

As execution proceeds,
\GMG evolves continuously.
When an operator is issued to a specific accelerator,
\GMG records its placement and issuing information and marks it as \textit{issued}.
When the operator finishes execution,
\GMG marks it as \textit{done} and releases the associated resource accounting.
\GMG is therefore not a static task description.
Instead,
it is a cluster-wide dynamic view that continuously evolves with execution and tracks the cluster state at operator granularity.

\textbf{Scheduling metadata.}
\sys maintains several key metadata for scheduling.
\emph{First},
when a graph template is registered,
\sys creates an empty operator estimator (\S\ref{sec:estimator}) for every operator in the template.
Each time an operator finishes execution,
its observed execution time is fed back to the corresponding estimator,
which is updated continuously over time.

\emph{Second},
\GMG has its own global timestamp to track the real world time,
with each node holding its timestamp to be \textit{issued} and \textit{done}.
Together with inter-operator dependencies,
these timestamps provide the starting point for temporal prediction of the global simulator.

\emph{Third},
each node stores the shapes of its output tensors.
These shapes are used to estimate the operator's execution time,
its HBM occupancy during execution,
and the latency of subsequent cross-accelerator tensor transfers.

\emph{Finally},
each node maintains checkpoint metadata that identifies which intermediate results are persistently retained at specific locations,
such as accelerator HBM.
It enables incremental scheduling by allowing new graph instances to be attached to retained intermediate states.


\subsection{Global Simulator}

The global simulator predicts how operator execution evolves across the cluster under different scheduling decisions.

\subsubsection{Transfer \& Operator Estimator}
\label{sec:estimator}

To support fast simulation, \sys maintains a set of lightweight linear estimators for tensor transfer and operator execution.

\textbf{Transfer estimator.}
For each accelerator pair, \sys maintains a linear estimator for transfer latency:
\begin{equation*}
t = \alpha \cdot size + c,
\end{equation*}
where $size$ is the tensor size, $\alpha$ reflects the inverse of effective bandwidth, and $c$ captures fixed transfer overhead.

\textbf{Operator estimator.}
For each pair of accelerator type and operator in graph template,
\sys maintains a linear estimator that predicts execution time from operator shape.
The input features are symbolic shape variables and shape-derived terms that capture both compute cost and tensor sizes.
For example,
for a batched matrix multiplication operator with shape parameters $(b,m,n,k)$,
where $b$ denotes batch size and $m$, $n$, and $k$ denote matrix dimensions,
\sys
uses
\begin{equation*}
t = \beta_1 (bmnk) + \beta_2 (bmk) + \beta_3 (bkn) + \beta_4 (bmn) + \beta_5,
\end{equation*}
where $bmnk$ captures the dominant compute cost,
$bmk$ and $bkn$ correspond to the sizes of the two input tensors,
$bmn$ corresponds to the size of the output tensor,
and $\beta_5$ captures shape-independent constant overhead.

The scheduler continuously collects performance feedback from the operator executor and incorporates them into the corresponding estimators online.
As more samples are fed back,
the estimators become more accurate.
These estimators provide the low-level latency input needed by the global simulator.

\subsubsection{Simulator Construction}

\begin{figure*}[t]
    \centering
    \includegraphics [width=0.9\linewidth]{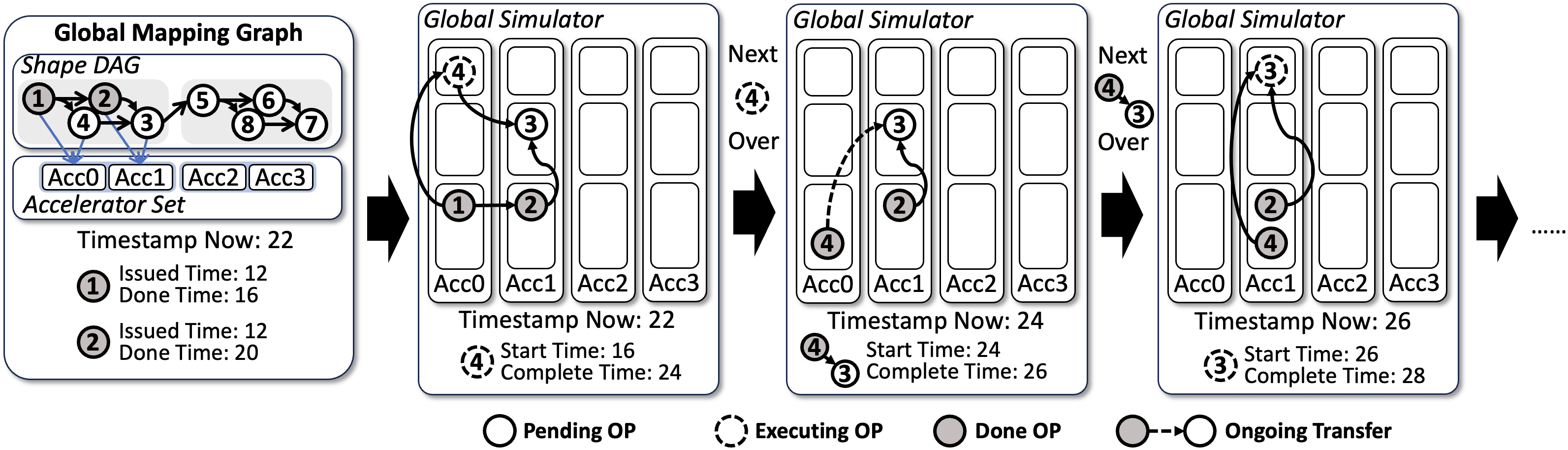}

    \vspace{-5pt}
    \caption{
        An example of the construction and simulation of the global simulator.
    }
    \label{simulation}
    \vspace{-10pt}
\end{figure*}






\GMG provides all information about the current cluster state,
which enables the construction of a global simulator.
Because \GMG records the mapping from operators to accelerators,
the system knows which \textit{issued} operators are assigned to each accelerator.
This allows the simulator to reconstruct the set of operators currently present in every accelerator.

The next step is to infer which operator is currently executing on each accelerator.
An \textit{issued} operator becomes ready once all of its predecessors are \textit{done} and their outputs are already available in the local HBM.
Its ready time is determined by the completion times of local predecessors,
and
by the arrival times of their transferred outputs for predecessors on other accelerators.
Each queue (\S\ref{sec:queue}) is designed to execute at most one operator at a time.
Therefore,
the operator currently in execution is the \textit{issued} operator with the earliest ready time.
Once the start time of an operator is known,
its completion time can be immediately predicted using the operator estimator.
This provides the timing information needed to drive subsequent timeline evolution.

Moreover,
the simulator can infer tensor residency in HBM.
For any ready operator,
all of its predecessor outputs must already be available on the accelerator to support the execution of the ready operator.
For the operator currently executing, its output must also have been allocated in HBM.
Based on these ready and executing operators,
the simulator can determine which tensors reside in HBM on each accelerator to infer current HBM usage.

Accordingly, given a \GMG snapshot at any time,
the system can reconstruct the contents of every accelerator queue at that instant,
including execution state and HBM occupancy.
This further shows that \GMG is equivalent to a consistent snapshot of the entire cluster at that moment.

\subsubsection{Event-Driven Simulation}




%
After reconstructing all accelerator queues from \GMG,
the simulator advances their states to predict how the cluster evolves over time.
It maintains a unified simulated timeline across all accelerators.
Specifically,
the simulator keeps track of all the running operators in the cluster,
together with their predicted completion times.

At each step,
it selects the earliest-complete operator as the completion event and advances simulated timeline to the complete-time.
The simulator then applies the cascading effects of this event.
When this operator finishes, one ready operator among those marked as \textit{issued} in the same queue starts executing.
For each newly started operator,
the simulator invokes the corresponding estimator to predict its completion time and inserts it into the operator running set.
The simulation then proceeds to the next event.

By iterating in this manner,
the simulator predicts, at operator granularity, both the cluster-wide execution timeline and the evolution of HBM usage over time.

\textbf{Summary.} As illustrated in Figure~\ref{simulation}, the global simulator operates in two steps.
It first reconstructs, from \GMG, the operators and dependencies on each accelerator and builds a corresponding queue for each accelerator in the simulator.
It then advances these queues event by event to infer how the cluster evolves over time.


This mechanism enables \sys to assess a candidate operator placement without actually deploying it on the real cluster.
\sys creates a copy of the current \GMG,
applies the placement to this copy,
and then constructs a global simulator based on it.
Using this simulator,
\sys derives a placement score from statistics on the simulated completion time of the relevant operators.
In essence,
it performs a what-if simulation.

\subsection{Scheduling Module}

\sys employs an independent scheduling module to orchestrate the execution of batches of \textit{unscheduled} graph instances.
The total number of \textit{issued} operators is limited to prevent excessive overhead of global simulation.
Therefore, it adopts a small-batch incremental scheduling strategy, and frequently invoked in response to both the insertion of new graph instances and runtime feedback.
In each scheduling round,
the scheduling module calls the incremental simulation-based scheduling algorithm
to collect the operators to be scheduled and generate their placement.


\subsubsection{Scheduling Frontiers}
To support incremental decisions,
\sys maintains two frontiers on \GMG.

The \emph{scheduled frontier} contains operators in state \textit{issued} or \textit{done} that have at least one \textit{unscheduled} successor;
these operators act as checkpoints that reside on accelerators for later scheduling.

The \emph{unscheduled frontier} contains operators in state \textit{unscheduled} whose predecessors are either absent or already \textit{issued} or \textit{done}.
All their predecessors lie on the scheduled frontier.
Together,
these frontiers define the current scheduling boundary on \GMG.

\begin{algorithm}[t]
\small
\caption{Incremental Simulation-Based Scheduling
}
\label{alg:scheduling}
\KwIn{Current \GMG, unscheduled frontier $F_u$, issued-operator budget $B$}
\KwOut{Operators to schedule and their placement}

\If{number of \textit{issued} operators in \GMG $\geq B$}{
    \Return
}

\If{$F_u$ is empty }{
    \Return
}

\medskip
\tcc{Batch Collection.}
$op \leftarrow$ highest-priority operator in $F_u$\;
$\mathcal{B} \leftarrow$ CollectConnectedGraph$(op, B)$\;

\medskip
\tcc{Simulation-Based Placement.}
bestScore $\leftarrow +\infty$\;
bestPlacement $\leftarrow \varnothing$\;

\ForEach{$p$ in CandidatePartitions \textbf{in parallel}}{
    $\GMG' \leftarrow$ Clone$(\GMG)$\;
    ApplyPlacement$(\GMG', \mathcal{B}, p)$\;
    
    sim $\leftarrow$ BuildGlobalSimulator$(\GMG')$\;
    feasible, score $\leftarrow$ Simulate$(sim, \mathcal{B})$\;
    
    \If{feasible = true}{
        RecordCandidateResult$(p,$ score$)$\;
    }
}

bestPlacement $\leftarrow$ BestRecordedCandidateResult$()$\;
\Return $\mathcal{B}$,bestPlacement

\end{algorithm}

\subsubsection{Incremental Simulation-Based Scheduling}

The input to each scheduling round is the current \GMG,
the unscheduled frontier,
and the budget on the number of \textit{issued} operators.
The output is a small batch of operators together with their placement.
The key constraint is responsiveness:
since \sys may be invoked frequently,
each round must remain lightweight enough that scheduling does not become a bottleneck.
To achieve this,
\sys combines batch collection and simulation-based placement,
as summarized in Algorithm~\ref{alg:scheduling}.

\textbf{Batch collection.}
\sys first selects the highest-priority operator from the unscheduled frontier,
and then follows its successors to collect a small connected batch of graph instances.
As serving requests often form dependency chains across graph instances,
this process naturally focuses each round on one high-priority request of a model instance.
The batch size is capped to keep scheduling overhead low.

\textbf{Simulation-based placement.}
Given the collected batch,
\sys evaluates candidate placements over a set of pre-defined accelerator partitions.
For each graph template,
the mapping of its operators to each candidate partition is also pre-defined.
This is possible because inference graphs are regularly decomposable:
once a partition is chosen,
the internal operator-to-accelerator mapping is fixed.
\sys applies each candidate placement to a cloned \GMG,
builds a global simulator,
and runs what-if simulation in parallel.
Placements that lead to OOM are discarded.
Among the remaining candidates,
\sys selects the one that minimizes the simulated average operator completion time.

After \sys calls the algorithm to get a placement,
\sys writes the chosen placement back to \GMG,
issues the corresponding operators to the operator executor,
marks them as \textit{issued},
and updates both frontiers.


This scheduling algorithm naturally captures both data locality and SLA priorities.
If early operators of a request run on one set of accelerators, and later ones run on another,
the required cross-accelerator tensor transfers are explicitly added to the simulated timeline.
This raises the predicted completion time of downstream operators.
Thus, placements that break locality are automatically disfavored.
The algorithm also protects high-priority SLAs:
because high-priority operators are more likely to be collected first from the unscheduled frontier,
under heavy load low-priority ones may not be collected at all.
Meanwhile,
the scheduler achieves fine-grained load balancing:
when a partition is heavily loaded,
its placement becomes less attractive or is ruled out during simulation.
Consequently,
\sys does not merely preempt low-priority work on a local accelerator;
it performs dynamic scheduling so the entire cluster prioritizes high-priority SLAs.

\subsection{Operator Executor}
Finally,
the operator executor conducts the inference.
We design tensor transfers as regular operators in the per-accelerator queue,
and their execution is gated by transfer arbitration.

\subsubsection{Per-Accelerator Operator Queue}

\label{sec:queue}

\begin{figure}[t]
    \centerline{\includegraphics[width=0.5\textwidth,]{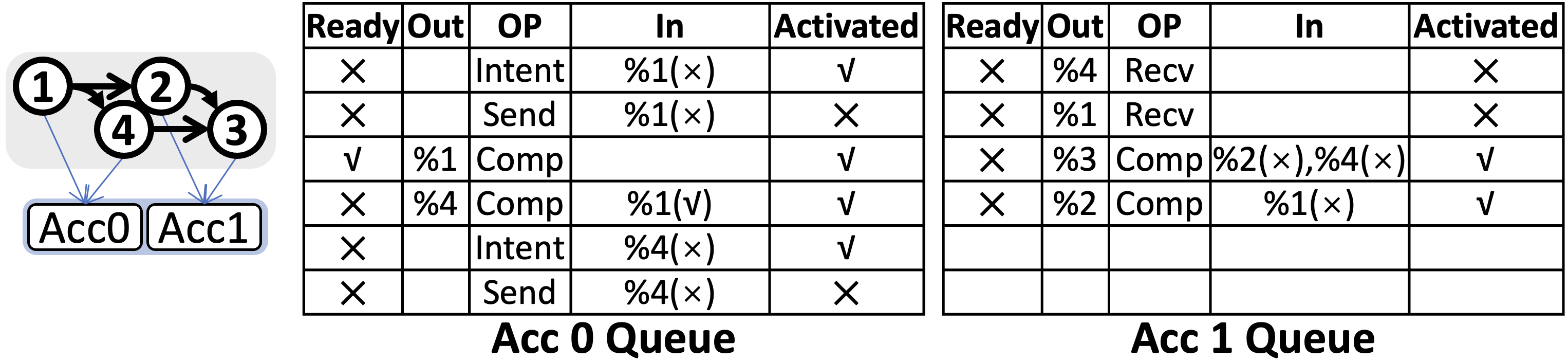}}
    
    \vspace{-10pt}
    \caption{
        An example of a computation graph partition and accelerator assignment.
        ``Compute'' refs to as ``Comp''. 
    }
    \label{queue}
    \vspace{-4mm}

\end{figure}

Each accelerator maintains a local operator queue as its device-side runtime.
The queue executes operators according to data dependencies rather than issuing order.
In this way,
a computation graph can be partitioned across multiple accelerators (as shown in Figure~\ref{queue}),
while each accelerator independently drives the operators placed on it.

Each queued operator records the readiness of its input operands and execution state.
When an operator is inserted into the queue, it is marked as \emph{ready} if it has no inputs;
otherwise,
its input slots are initialized as \emph{not ready} and the operator remains blocked.
Another execution state is \emph{activated},
which is used to determine whether an operation is permitted to execute.

The accelerator keeps listening the ready set.
Whenever the device is available,
it selects one \textit{ready} and \textit{activated} operator and executes it.
Each queue runs at most one operator at a time.
After an operator completes,
its output will be propagated to dependent operators.
For each successor,
the corresponding input slot is marked ready;
once all of its input slots are ready,
the successor becomes ready as well.

\subsubsection{Tensor Transfer Arbitration}


Graph partitioning in \sys creates remote dependencies,
where an operator on one accelerator consumes a tensor produced on another.
Such dependencies require explicit tensor movement across accelerators.

A naive design would let queues invoke point-to-point primitives as soon as source tensors become available.
However, these primitives are typically blocking, non-preemptible, and mismatched send/recv ordering across devices can easily create circular waits.
Therefore, \sys does not allow queues to autonomously start cross-accelerator transfers.

Instead, \sys coordinates such transfers through arbitration-managed transfers and an arbitration algorithm.

\medskip
\textbf{Cross-accelerator transfers.}
Each cross-accelerator transfer is decomposed into three operators: \textit{Intent}, \textit{Recv}, and \textit{Send}.
When the source tensor becomes available,
the \textit{Intent} operator notifies \sys of the pending transfer request.
Once the transfer is admitted,
\sys activates \textit{Recv} on the destination accelerator to allocate the receive buffer.
After the buffer is successfully prepared,
\textit{Recv} triggers \textit{Send} on the source accelerator to initiate the actual data movement.
When the reception completes,
\textit{Recv} reports completion back to \sys.

We enforce this Recv-before-Send ordering, as destination-side buffer allocation may temporarily fail and require retries.
By delaying \textit{Send} until \textit{Recv} succeeds,
the sender only invokes the communication primitive when the destination is ready.



\medskip
\textbf{Arbitration algorithm.}
To coordinate tensor transfers at cluster scope,
\sys maintains a communication graph,
separate from \GMG,
whose vertices represent accelerators and whose edges represent pending transfer intents.

When an \textit{Intent} operator executes,
\sys inserts the corresponding edge into the graph.
Because pending transfers may share the same source, share the same destination,
or compete for the same communication link,
\sys cannot activate every intent immediately.
Instead,
it selects a conflict-free set of transfers that can proceed concurrently.

For each selected edge,
\sys starts the transfer procedure by activating the destination-side \textit{Recv}.
When receiving a completion feedback,
\sys removes the corresponding edge and releases the associated resources.

This arbitration is event-driven.
It is triggered by the arrival of a new \textit{Intent} and the completion feedback of an existing transfer.
At each trigger,
\sys re-examines the communication graph and selects a new set of activatable edges.
In this way,
\sys avoids deadlocks caused by accelerator-local negotiation while allowing non-conflicting point-to-point transfers to proceed in parallel.




\section{Implementation}




We implement \sys in three layers:
the serving application interface,
the \sys master,
and the operator executor.

At the interface level,
we integrate \sys into PyTorch as a custom backend.
Specifically,
when a module is executed for the first time,
we use TorchDynamo~\cite{dynamo} to capture its computation graph.
The captured graph is then registered with \sys as a graph template,
and instantiated.
In this way,
\sys can be integrated into existing PyTorch-based serving applications with minimal changes to application code.

The \sys master runs on the host CPU and implements cluster-wide scheduling and transfer coordination.
Its operator executor is implemented 
on top of LibTorch~\cite{libtorch},
and runs on the accelerator side to execute operators and carry out tensor transfers.

\section{Evaluation}

\subsection{Experimental Setup}
\label{sec:setup}

\textbf{Hardware.}
We deploy and evaluate \sys on a single multi-GPU server.
The server is equipped with two Intel Xeon Platinum 8468 CPUs~\cite{intel8468} (96 CPU cores in total), 1600~GB of host memory,
and eight NVIDIA A16 GPUs~\cite{a16}, each with 64~GB of GPU memory.
We use GPU IDs 0 through 7 throughout the evaluation.

\textbf{Baselines.}
We compare \sys against an \textit{Static-Online} baseline and three GPU-sharing baselines. 
\begin{itemize}
    \item \textit{Static-Online}:
    Four model instances serve the four online traces separately,
    each pinned to a fixed GPU pair: (0,1), (2,3), (4,5), and (6,7).

    \item \textit{Subset-Share}: Priority-aware sharing~\cite{antman} on a fixed GPU subset. It launches an additional low-priority offline instance on a fixed 4-GPU subset (i.e., GPUs (0,1,2,3)).
    It opportunistically delays low-priority GPU operators to expose idle slices for other.

    \item \textit{Dual-Share}: Interference-aware sharing~\cite{orion}. It launches two additional low-priority offline instances on two 4-GPU subsets
    (i.e., GPUs (0,1,2,3) and (4,5,6,7)).
    It spreads offline requests across multiple fixed GPU subsets to reduce interference.

    \item \textit{Full-Share}: Full-server fixed-scope sharing.
    It launches only one additional low-priority offline instance across all eight GPUs.
    This single offline workload attempts to exploit idle slices over all the GPUs.
\end{itemize}

We implement these baselines ourselves to capture the key designs for prior works,
because existing GPU sharing systems are not always publicly available and often differ substantially in workload assumptions and runtime mechanisms

\textbf{Metrics.}
We use the following metrics.
\begin{itemize}
    \item \emph{Request throughput}: average number of requests processed per second during execution.
    \item \emph{Token throughput}: average number of output tokens generated per second.
    \item \emph{TTFT} (time to first token): the elapsed time from the moment a request arrives to the moment the first output token is produced; it therefore includes both queuing delay and prefill computation time.
    \item \emph{SLO attainment}:
    the p90 TTFT of the \textit{Static-Online} as the threshold: it is the fraction of online requests whose TTFT is below this threshold.
    \item \emph{GPU Utilization}: the fraction of a time window occupied by operator-execution slices on the GPU.
    \item \emph{Simulation throughput}: The number of operators simulated per millisecond in the simulator.
\end{itemize}

\textbf{Inference workloads.}
We adapt the official Llama3-8B model~\cite{llama3github} to run with \sys.
The online inference workload is driven by four representative one-day traces introduced in \S\ref{sec:trace},
which emulate the dynamic request patterns in deployment.
We replay the trace segment from 19:00 to 19:30, which is a representative interval with both busy and idle periods.
For offline inference,
we issue one request per second,
each with an input length of 2048 tokens and a fixed output length of 15 tokens.
This setting represents a practical lightweight offline inference workload in production.

\textbf{Operator granularity.}
A single inference/training graph execution typically involves thousands to tens of thousands of ATen ops~\cite{aten}.
If each ATen op were treated as a scheduling unit,
the scheduling overhead would be prohibitive.
Therefore,
we coarsen the scheduling granularity by merging the forward or backward computation of each layer into one operator,
and use layers as the scheduling unit in \sys.

\begin{figure*}[t]
\centering
\includegraphics[width=0.7\linewidth]{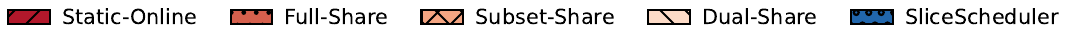} \\
\vspace{-0.2cm} 
\subfloat[Request throughput.]{\includegraphics[width=0.24\linewidth]{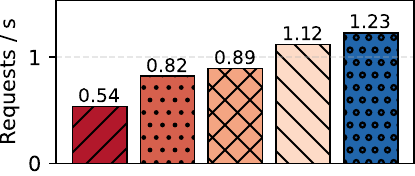}}
\hfill
\subfloat[Token throughput.]{\includegraphics[width=0.24\linewidth]{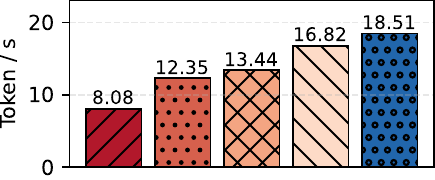}}
\hfill
\subfloat[SLO attainment.]{\includegraphics[width=0.24\linewidth]{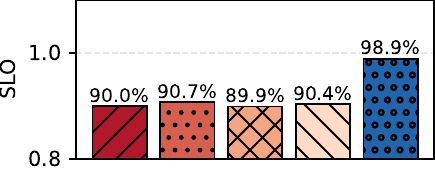}}
\hfill
\subfloat[GPU utilization.]{\includegraphics[width=0.24\linewidth]{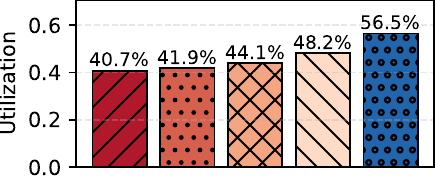}}

\caption{
Performance comparison of \sys and baselines.
}
\label{onlineoffline}

    \vspace{-10pt}
\end{figure*}

\subsection{Online/Offline Co-location Throughput}
\medskip

To evaluate the effectiveness of dynamic scheduling in exploiting idle slices,
we assess \sys in an online/offline co-location setting.
In this setting,
\sys runs with both online and offline requests admitted to the system.
The online inference workload generates idle slices,
while the offline workload attempts to fill these slices and convert them into useful throughput.
We assign a higher execution priority to operators from the online workload than to those from the offline workload.


Figure~\ref{onlineoffline} shows the results.
Overall,
\sys improves SLO attainment by about 9\% over all baselines,
increases request throughput by 9\%--129\%,
improves token throughput by 10\%--129\%,
and raises GPU utilization by 17\%--38\%.
GPU utilization is generally low in this experiment because the selected trace segment is relatively light, and the admitted offline workload is intentionally lightweight as well.
Nevertheless,
the observed SLO improvement shows that the trace still contains periods of substantial contention. We next demonstrate that, by admitting additional offline workload into these fragmented slack periods, \sys can further improve utilization while preserving online performance.


Importantly,
these throughput gains do not come at the expense of online service quality.
Compared with all baselines,
This improvement comes from breaking container-level isolation and enabling finer-grained load balancing across the whole cluster.
While the GPU-sharing baselines already incorporate priority-aware~\cite{antman} co-location,
their fixed resource scope still limits their ability to preserve online performance under fragmented idle slices.
In contrast,
\sys combines priority-aware scheduling with cluster-wide operator-level remapping,
allowing it to better protect high-priority online requests even in the presence of offline workloads.

At the same time,
\sys achieves higher throughput and GPU utilization than all baselines,
with the gains mainly coming from the throughput contributed by offline work and from opportunistic inserting offline requests into idle slices.

Compared with \textit{Full-Share},
\sys improves request throughput by 49\%.
The main reason is that,
although \textit{Full-Share} also allows an offline workload to span the whole machine,
\sys additionally performs automatic load balancing,
which increases the duration of overlapping idle slices across multiple GPUs and thus improves the effective utilization of multi-GPU offline execution.

Compared with \textit{Subset-Share} and \textit{Dual-Share},
\sys improves throughput by about 9\%--37\%.
The reason is that existing GPU sharing approaches~\cite{antman, orion} typically co-located low-priority work on a fixed GPU scope,
whereas \sys performs cluster-wide dynamic scheduling and can therefore exploit idle slices from a much larger resource pool.

\subsection{Overhead of \sys}

\medskip
\begin{table}[t]

\small
    \centering
    \caption{Execution time for running 1,000 prefills. Relative overhead is computed against \textit{Direct Execution}.
    }
    
    \begin{tabular}{lcc}
    \hline
    \textbf{Execution Mode} & \textbf{Time (s)} & \textbf{Relative Overhead} \\ 
    \hline
    \textit{Direct Execution} & 23.39 & -  \\
    \textit{\sys} & 23.37 & 0.0\% \\
    \textit{Eager Execution} & 25.27 & 8\% \\
    \hline
    \end{tabular}
    \label{overhead}
\end{table}




We evaluate the overhead introduced by the \sys.
To minimize the actual computation time and thereby make system overhead more visible,
we run 1000 prefill executions with a context length of 1 on a single GPU and measure the total execution time.

We consider three execution modes.
(i) \textit{Direct Execution} executes the inference graph directly and controls the GPU without any intermediate system layer; it therefore serves as the overhead baseline.
(ii) \textit{\sys} executes the same inference graph through the \sys scheduling layer on a single GPU.
(iii) \textit{Eager Execution} runs the workload directly from the user program.

Table~\ref{overhead} shows the results.
The execution time of \sys is almost identical to that of \textit{Direct Execution},
indicating that the operator scheduling layer adds negligible runtime overhead beyond the direct execution path.
For reference, \sys is 8\% faster than \textit{Eager Execution},
mainly because it avoids Python interpreter overhead in eager mode~\cite{eagermode}.
Overall,
these results show that \sys remains close to the lower-level execution path while adding little extra cost of its own.


\subsection{Accuracy and Throughput of Global Simulator}


To evaluate whether our simulator can effectively support dynamic scheduling,
we focus on simulation accuracy and throughput.
To the best of our knowledge,
we are the first fast cluster-wide simulator designed for dynamic scheduling in multi-tenant AI clusters.
Therefore, we adopt SimAI~\cite{simai} as our baseline,
which is the most closely-related state-of-the-art solution.

Since \textit{SimAI} is primarily designed for training workloads,
we use the Llama3-8B training computation graph to evaluate both \textit{SimAI} and our global simulator.
We first statically map the training graph onto two GPUs with a pipeline parallelism configuration of $PP=2$,
and run it for 256 iterations to allow the estimators to collect runtime data, including operator execution time, tensor transfer time and the end-to-end per-training time. Using these collected data,
we construct an empty global simulator, putting the same statically mapped training graph into the simulator’s execution queue.
When the simulator executes, we collect the simulation throughput and the simulated execution time of each training iteration.

\subsubsection{\sys estimator accuracy.}

\begin{figure}[t]
\centering
\subfloat[Execution-time samples and predictions for representative operators.]
{\includegraphics[width=0.485\linewidth]{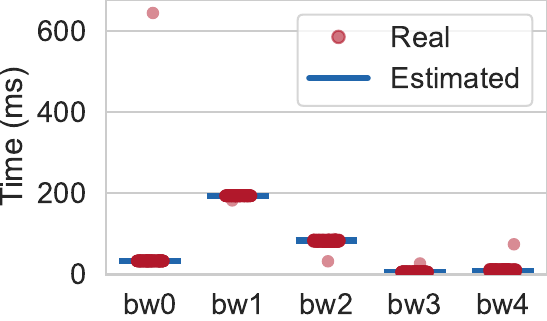}}
\hfill
\subfloat[Transfer-bandwidth samples and predictions for representative tensor sizes.]{\includegraphics[width=0.485\linewidth]{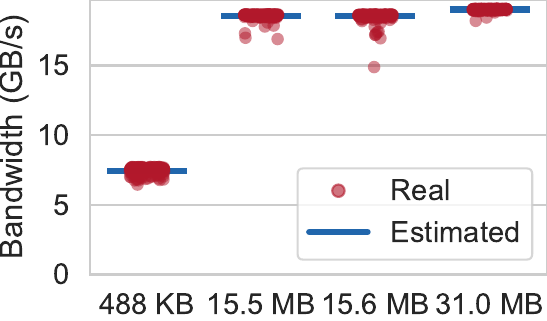}}

\caption{
Prediction accuracy of the operator and transfer estimators.
}
\vspace{-5mm}
\label{estimator}
\end{figure}

Figure~\ref{estimator} compares real and estimated execution times for backward operators in the upper half layers,
as well as real and estimated transfer latencies for several representative tensors.

We observe that operator execution time is highly stable across iterations,
and tensor-transfer bandwidth is also relatively stable.
The coefficient of variation is 0.62\% for operator samples and 1.63\% for transfer samples.
As a result,
the predictions produced by these estimators are strongly correlated with the measured execution time.
The mean absolute percentage error of the operator and transfer estimators is 0.32\% and 0.92\%, respectively.
These results validate the use of linear regression to predict both operator execution time and tensor-transfer cost.

\subsubsection{End-to-end accuracy.}


\begin{table}[t]
\small
\centering
\caption{Average training time per one iteration of Llama3-8B on dual GPUs and simulation error rate.}

\label{simulation_e2e}
\begin{tabular}{lcc}
\hline
\textbf{Method} & \textbf{Time (s)} & \textbf{Error Rate} \\
\hline
Real & 1.910 & - \\
Ours & 1.945 & 1.83\% \\
\textit{SimAI} & 1.517 & 20.58\% \\
\hline
\end{tabular}
\end{table}



Table~\ref{simulation_e2e} summarizes the actual and predicted end-to-end execution time of one training iteration.

Our global simulator achieves an average error rate of only 1.83\%,
showing that operator-granularity simulation can accurately characterize the cluster-wide execution of operators.
In comparison, \textit{SimAI} has an average error rate of 20.58\%.
The reason is that it does not use estimators that continuously learn from runtime feedback,
while \sys incrementally refines both operator and transfer performance estimators during execution.

\subsubsection{Simulation throughput.}

\begin{figure}[t]
\centering
\subfloat[Global simulator]{\includegraphics[width=0.485\linewidth]{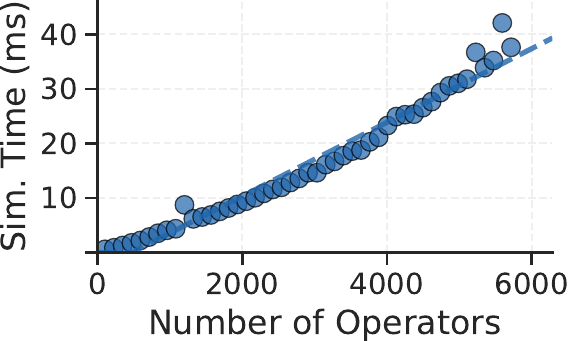}
}
\hfill
\subfloat[\textit{SimAI}]{\includegraphics[width=0.485\linewidth]{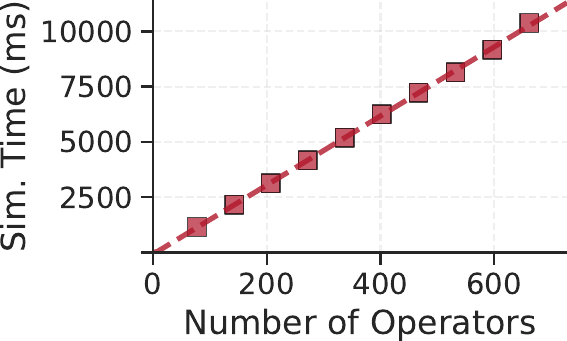}}

\caption{
Simulation time versus simulated operators for our global simulator and \textit{SimAI}.
"Sim." refers to "Simulation".
}
\label{simulatiton_time}
\end{figure}


We measured the simulation throughput of our global simulator.


For the global simulator,
we accumulate different numbers of un-simulated operators and then run the simulation to measure the simulation time,
from which we derive the simulation throughput.
As shown in Figure~\ref{simulatiton_time}~(a),
the simulator processes on average $\sim$176.8 operators per millisecond.

For \textit{SimAI},
we also simulate models of different sizes to obtain the equivalent number of simulated operators and the corresponding simulation time.
As shown in Figure~\ref{simulatiton_time}~(b), \textit{SimAI} processes on average $\sim$0.065 operators per millisecond.
Thus,
the global simulator achieves about four orders of magnitude higher simulation throughput than \textit{SimAI}.
This improvement stems from the different goals of the two simulators: our simulator de-emphasizes traffic scheduling,
which significantly reduces simulation complexity.

\subsection{Scalability of Dynamic Scheduling}







To understand the limits of simulation-based dynamic scheduling, we analyze the scalability of \sys.

The benefit of \sys comes from prediction horizon --- 
a longer prediction window helps identify fine-grained load imbalance and improves placement quality.
However,
it also increases the number of operators to simulate and thus the cost of each scheduling round.
If scheduling takes too long,
the cluster keeps executing while the centralized scheduler is still running what-if simulation,
making scheduling itself a bottleneck.
\sys must therefore balance decision quality against scheduling timeliness.

We focus on four dimensions:
\begin{itemize}[leftmargin=*]
\item \textbf{Operator granularity} ($g$): The average execution time of one scheduling unit on an accelerator.
Finer granularity means more operators must be scheduled and simulated within the same amount of execution time.

\item \textbf{Scheduling volume} ($v$): The total execution time of the operators selected in one scheduling round.

\item \textbf{Prediction horizon} ($l$): The future time window simulated in one scheduling round, normalized by the scheduling volume.
A larger horizon improves foresight but increases simulation cost.

\item \textbf{Scheduling timeliness} ($t$): The ratio between the number of operators executed by the cluster during the scheduling time, and the number of operators scheduled in one round.
Smaller is better.
When this ratio approaches or exceeds 1,
the scheduler becomes a bottleneck and the benefit of dynamic scheduling quickly diminishes.
\end{itemize}

Let the number of accelerators be $N$, the simulation throughput be $s$, the operator granularity be $g$, the prediction horizon be $l$, the scheduling volume be $v$, and the scheduling timeliness be $t$.
Therefore, scheduling timeliness is
$t=\tfrac{Nl+1}{sg}$.



In our experiment setup,
we merge every 8 of the 32 layers of Llama3-8B into one operator.
Given an average decode latency of $\sim$42.3~ms,
the resulting operator granularity is $g=10.575$.
Using the measured averaged simulation throughput $s=176.8$,
a prediction horizon of $l=3$, and a target timeliness of $t=0.1$, the above analysis yields $N \approx 62$.
This suggests
\sys can scale to clusters with up to about 64 GPUs.

In practice, we tune the scheduling volume $v$ and prediction horizon $l$ offline.
The average operator granularity is controlled by empirically grouping LLM layers into scheduling units.
Adding the number of accelerators,
\sys can estimate the number of operators simulated in each scheduling round.
This estimate is then used to cap the number of \textit{issued} operators maintained in \GMG,
ensuring \sys remains practical for small to medium-scale clusters.
\section{Related Work}






\textbf{Job scheduling and elasticity.}
Cluster schedulers such as Kubernetes~\cite{k8s}, Ray~\cite{ray}, and Firmament~\cite{firmament} schedule application processes or containers in the cluster.
Such systems operate at the granularity of processes rather than the code directly executed on GPUs, making the scheduling unit too heavyweight to exploit large numbers of transient idle slices.
To handle changing workload demand,
prior work introduces elasticity mechanisms~\cite{torchelastic,lyra,shockwave,easyscale,bian2021online,pollux}
to re-assign resources to the job.
However, due to the heavyweight of the scheduling payload,
elasticity incurs substantial overhead.
As a result, these approaches are ill-suited to the highly transient and fine-grained idle slices targeted by \sys.

\textbf{GPU sharing.}
A large body of work improves utilization by allowing multiple jobs to share one GPU~\cite{mig,mps,orion,gpuvirt,antman,lithos}.
However,
they primarily operate within the scope of a single GPU or a fixed GPU subset,
and do not fully exploit the complementary idle slices distributed across GPUs at cluster scale.
As a result,
handling fragmented idle slices across the cluster requires global scheduling.
\sys complements GPU-sharing approaches by introducing operator-level, cluster-wide scheduling to reclaim such idle slices.

\textbf{Simulation.}
Prior simulators mainly target either local hardware behavior or large-scale distributed execution.
Micro-architectural simulators~\cite{gem5,gpgpusim} provide cycle-level modeling of processors and GPU kernels,
but they focus on intra-operator behavior and are far too slow for cluster-scale scheduling.
At a higher level,
distributed and network simulators~\cite{astrasim,simai} model AI workloads on large clusters.
However, their simulation time can still be on the order of hours for realistic workloads, and they do not explicitly reason about HBM occupancy or out-of-memory conditions.
In contrast,
the \sys global simulator is designed for dynamic scheduling: it performs fast operator-level simulation while explicitly tracking both execution time and HBM usage.

\textbf{LLM Serving.}
They typically combine local and global scheduling.
Local schedulers~\cite{pagedattention,pastfuturescheduler} manage requests within a fixed model deployment to meet SLA targets, while global schedulers~\cite{nvidiadynamo} distribute requests across deployments for load balancing.
However, global schedulers usually operate at coarse granularity and rely on low-frequency utilization signals.
\sys bridges these two levels: it combines cluster-wide coordination with fine-grained execution control, and directly uses executable operators, rather than requests or containers, as the scheduling unit.

\textbf{Graph-based scheduling.}
MapReduce~\cite{mapreduce} and Spark~\cite{spark} pioneered graph-based execution by representing applications as computation DAGs and scheduling tasks according to data dependencies.
Their abstractions are effective for coarse-grained data analytics workloads,
but are not designed for fine-grained GPU execution in AI clusters.
\sys similarly leverages graph structure,
but uses GPU executable operators as the scheduling unit and explicitly reasons about accelerator placement, execution timing, and HBM usage.

\section{Conclusion}

In this paper,
we present \sys,
an operator-level scheduling system for reclaiming fine-grained idle slices in multi-tenant AI clusters.
By combining the Global Mapping Graph,
an event-driven global simulator,
simulation-based scheduling,
and an operator executor,
\sys improves cluster utilization while preserving online SLAs.
Results show \sys achieves higher throughput and utilization than container-level and GPU-sharing baselines,
while introducing negligible runtime overhead.

\bibliographystyle{ACM-Reference-Format}
\bibliography{ref}

\end{document}